\documentclass[pra,aps,amsfonts,psfig]{revtex4}
\usepackage{psfig}
\usepackage{graphicx}

\def\textbf#1{{\bf #1}}

\def\eea{\end{array}}
\def\bea{\begin{array}}
\def\tr{{\rm Tr}}
\def\id{{\rm I}}

\def\>{\rangle}
\def\<{\langle}
\def\be{\begin{equation}}
\def\ee{\end{equation}}
\def\ben{\begin{eqnarray}}
\def\een{\end{eqnarray}}
\def\beng{\begin{eqnarray*}}
\def\eeng{\end{eqnarray*}}
\newcommand{\bei}{\begin{itemize}}
\newcommand{\eei}{\end{itemize}}
\newcommand{\bee}{\begin{enumerate}}
\newcommand{\eee}{\end{enumerate}}

\newtheorem{theorem}{Theorem}
\newtheorem{fact}{Fact}
\newtheorem{lemma}{Lemma}[section]

\def\pcal{{\cal P}}
\def\blacksquare{\vrule height 4pt width 3pt depth2pt}
\def\ot{\otimes}

\begin{document}

\title{Bounds on localisable information via semidefinite programming}
\author{Barbara Synak$^{(1)}$, Karol Horodecki$^{(2)}$, Micha\l{} Horodecki$^{(1)}$}

\affiliation{$^{(1)}$Institute of Theoretical Physics and Astrophysics,
University of Gda\'nsk, Poland}
\affiliation{$^{(2)}$Department of Mathematics, Physics and Computer Science
University of Gda\'nsk, Poland}

\begin{abstract}
We investigate so-called {\it localisable information}  of  bipartite states and a parallel notion of {\it information deficit}.
Localisable information is defined as the amount of information that can be concentrated by 
means of  classical communication and local operations where only maximally mixed states 
can  be added for free. The information deficit is defined as difference between total 
information contents of the state  and localisable information. We consider a larger class of
 operations: the so called PPT operations, which in addition preserve maximally mixed state 
(PPT-PMM operations).  We formulate the related optimization problem as sedmidefnite 
program with suitable constraints. We then provide bound for fidelity of transition of a given 
state into product pure state on Hilbert space of dimension $d$. This allows to obtain general 
upper bound for localisable information  (and also for information deficit). We calculated the
bounds exactly for Werner states and isotropic states in any dimension. Surprisingly it turns 
out that related bounds for information deficit are equal to relative entropy of entanglement 
(in the case of Werner states - regularized one). 
We compare the upper bounds with lower bounds based on simple protocol of localisation of information.
 \end{abstract}
\maketitle

\section{Introduction}
In recent development \cite{OHHH2001,nlocc} an idea of {\it localizing 
information (or concentrating)} in paradigm of distant laboratories was devised. It originated from 
the concept of drawing thermodynamical work form heat bath and a source of negentropy 
(see e.g. \cite{Vedral1999,Scully-negentropy}). Namely, using one qubit in pure 
state, one can draw $kT\ln 2$ of work \cite{AHHH} from heat bath of temperature T .
More generally, using $n$ qubit state $\rho$ one can draw $n-S(\rho)$ of work.
(We neglect the obvious factor $\ln kT$, counting work in bits).
In \cite{OHHH2001} this idea was applied to the distant labs paradigm.
There are distant parties, who share some $n$ qubit quantum state,
and have local heat baths of temperature $T$. If the parties can communicate 
quantum information, then they can use the shared state to draw  $n - S(\rho)$ 
bits work. This can be achieved, 
by sending the whole subsystem to one party. The party can then draw work from 
local heat bath by use of the total state.
However, if they can only use local operations and classical communication (LOCC), 
then they usually will not be able to draw such amount of work.
Indeed, if they try to send all subsystems to one party, the state 
will be decohered, due to transmission via classical channel. 
Thus all quantum correlations will disappear, which will result in increase of 
entropy of the state to some value $S'>S$. Thus we will observe a difference 
between total information $n-S$ and information localisable by LOCC. 
The difference is called {\it quantum deficit} and denoted by $\Delta$.
Since it represents the information that 
must be destroyed during travel through classical channel, it reports 
quantumness of correlations of a state.  In this way tracing what is local, 
we can also understand what is non-local. 

The basic problem arises:
{\it Given a quantum compound state, how much 
information can be localized by LOCC?}.  Or, equivalently,
{\it How large is quantum deficit for a given state?}
For pure state the answer is known \cite{nlocc}: the amount of information 
that cannot be localized is precisely entanglement of formation of the 
state \cite{BDSW1996}, given by entropy of subsystem. However, for mixed states
even separable states can have nonzero deficit. Thus the deficit can account for 
quantumness that is not covered by entanglement. One could also expect that 
deficit is the measure of all quantumness of correlations, so that 
reasonable entanglement measures should not exceed it. In any case, it is important 
to evaluate deficit for different states.

In \cite{compl} the problem of localising information into subsystem was translated 
into a problem of distilling pure local states.
In this paper, basing on this concept,  we provide general upper bound for the localisable information,
 which gives in turn  
lower bound for information deficit. We calculate the bound for 
 symmetric states, like Werner states \cite{Werner1989} and 
isotropic states \cite{reduction}. We use method of semidefinite programming 
following Rains approach to the problem of entanglement distillation \cite{Rains2001}. 
Though our problem is quite opposite: instead of singlets, we want 
to draw pure product states, the technique can be still applied, and the bounds 
we obtain share some features of Rains bound for distillable entanglement.
Even more suprisingly, the bounds for information deficit obtained
for Werner  and isotropic states  are just {\it equal} to Rains bounds for
those states, which in turn are euqal to
relative entropy of entanglement (regularized in the case of Werner states).
We also present lower bounds, obtained by some specific protocols 
of localizing information. 



\section{Definitions}

In this section we will set some definitions. A quantum operation is completely 
positive trace preserving (CPTP) map.   In entanglement theory 
one distinguishes among others LOCC operations \cite{BDSW1996}(local operations 
and classical communication) - these are operations that can be 
applied to a state of compound system by using any local measurements, local operations
and classical communication. For bipartite systems there is also $PPT$ class. There belong 
operations $\Lambda$ for which $\Gamma\Lambda\Gamma$ is still an legitimate operation (where 
partial transpose $\Gamma$ is given by $\Gamma=I\otimes T$, 
with $T$ being matrix transposition.

If we are interested in localizing information, we have to find a way to 
count the information. In the above classes,   one can add pure local 
states for free. Thus we have to restrict the classes, to be able 
to trace the information flow. Consequently, following  \cite{nlocc},
we consider the class NLOCC (noisy LOCC), which differs from original LOCC 
in one detail: one cannot add arbitrary  local ancilla. 
Only maximally mixed ancillas can be added. \\
We will also consider operations, which preserves maximally mixed state and  are PPT simultaneously.
 They will be called {\bf PPT-PMM} operations.

One can ask whether adding pure ancillas can help in case when we can control amount of  
information given by them and subtract it from final result. In \cite{huge} it will be shown, that in
general, such catalysis is not useful.

In any paradigm of manipulating states by operations the basic 
notion if {\it rate of transition}. 
Given a class of operations, 
one can ask, at what rate it is possible to transform  state $\rho$ into $\sigma$, 
given large $n$ independent copies of $\rho$, 
\be 
\rho^{\ot n} \to \sigma^{\ot m} 
\label{transition} 
\ee 
The above means that acting on $\rho^{\ot n}$ by one of allowed operations, 
we get some state $\sigma_m$ that is close to the desired state $ \sigma^{\ot m}$ in trace norm
for large $n$. The optimal rate $R(\rho\to \sigma)$ is defined as 
limsup of $m\over n$, where we take limit of large $n$. 

In our case, the target state is local pure state. The localisable information $I_l$ 
of $\rho$ is the amount of such local pure qubits obtained from $\rho$ per input pair, 
by NLOCC. Throughout the paper, we will work with converting states into {\it pairs of qubits},
as this is more convenient. 
We will use  to denote twice rate of transition in specific protocols, 
as well as, depending on the context, in maximal  protocol. 
Thus given $n$ copies of initial state $\rho$ 
and $m$ copies of final two qubit pair $|0\>|0\>$, 
by $r$ we mean $2m/n$. Sometimes we will write $r_\pcal$ which 
denotes rate in protocol $\pcal$. Optimal $r$ over all protocols is of course $I_l$. 

The information deficit $\Delta$ is the difference between 
 total information and localisable information 
\be
\Delta(\rho)=n-S(\rho)- I_l(\rho)
\label{delta}
\ee

We will also need definitions of entanglement measures.

\bee
\item{\it Entanglement of distillation \cite{BDSW1996} $E_D$} is a maximal number 
of singlets per copy distillable by LOCC operations from the 
state $\varrho$  
in asymptotic regime of $n \rightarrow \infty$ copies.
\item {\it Entanglement cost $E_c$}  is minimal number of 
singlets per copy needed to create a state $\varrho$ by LOCC operations in asymptotic 
regime of $n\rightarrow \infty$ copies.
\item {\it Relative entropy of entanglement $E_R$}
 is defined as follows:
 \be
 E_R(\varrho)=\inf_{\sigma\in SEP}S(\varrho|\sigma)
 \ee
where $S(\varrho|\sigma)=tr\varrho\log_2\varrho-tr\varrho\log_2\sigma$ is relative 
entropy and the infimum is taken  over all separable states $\sigma$.
\item {\it Regularized relative entropy of entanglement $E^\infty_R$}
is given by the formula:
\be 
E^\infty_R(\varrho)= \lim_{n\rightarrow \infty} \frac{E(\varrho^{\otimes n})}{n}
\ee
\eee

\subsection{Werner and isotropic state} 
Here we will recall two well known families of states.
Let us consider states, which do not change if subjected to the 
same unitary transformation to both subsystems:
\be 
\varrho=U\otimes U \varrho U^{\dagger}\otimes U^{\dagger} \quad \textrm{for any unitary U}
\label{ww}
\ee
Such state (called Werner state) must be of the following form \cite{Werner1989}:
\ben
\varrho_{W}=\frac{1}{d^{2}+d\beta}(I+\beta V),\quad -1\leq \beta \leq 1
\label{wer1}
\een
where V  a unitary flip operator V on $d\otimes d$ system difined by  
$V\phi\otimes\varphi=\varphi\otimes\phi$.
Other form for $\varrho_{W}$ is 
\ben
\varrho_{W}=p \frac{P_{A}}{N_{A}}+(1-p)\frac{P_{S}}{N_{S}} 
\label{wer2}
\een
where $P_{S}(P_{A})$ is projector onto symmetric (antisymmetric) subspace of total space,
$N_{A}=(d^{2}-d)/2$ ($N_{S}=(d^{2}+d)/2$) is the dimension of the antisymmetric (symmetric)  
subspace.

There are states, which are invariant under $U\otimes U^{*}$ transformation \cite{reduction}.
 The state, called isotropic
state  are of the following form:
\be
\varrho_{iso}=\lambda P_{+} + \frac{1-\lambda}{d^{2}}I \quad \quad \lambda \in [-\frac{1}{d},1]
\ee
where $P_{+}$ is maximally entangled state and I is the identity state.

\section{Bounds on the fidelity of transitions}
\label{sec:fidelity-loc}
Given a quantum mixed state $\rho$ one may ask how much of the information
it contains which can be concentrated to a local form \cite{OHHH2001}. 
In other words how much pairs of a pure {\it product} 
states $P_{00}$ one can achieve from $\rho$ per copy of 
$\rho$  in asymptotic regime under noisy local operations and classical communication (NLOCC)
\cite{nlocc}. The class of NLOCC maps is rather difficult to deal with. Then similarly as 
in entanglement theory, it is more convenient to consider some larger class that has clear 
mathematical characterization. In \cite{Rains1999} a class of PPT maps was introduced which is 
larger than LOCC. 
In our case, the analogous larger class will be PPT-PMM maps. If we are able to get 
upper bound for the
rate of distillation of pure product states  with this new class it will be also upper bound for
rate achievable by NLOCC maps. Our analysis of the rate under PPT-PMM  maps , will be analogous to 
the analysis of distillation of entanglement in \cite{Rains2001}.
Having fixed the rate of conversion from $\rho$ to $P_{00}$ we evaluate the fidelity of 
conversion i.e. the overlap of the current output with desired output. If the fidelity 
can approach 1 in limit of many input copies, the rate is attainable.  

Let us then fix the rate $r$ which means, that for
$n$ input copies of a given state we will obtain $ m = nr/2 $ output copies.
The $m$ pairs are in a final joint state $\rho ' = \Lambda (\rho^{\otimes n}) $, where $\Lambda$
is an operation of conversion. In the following we will assume, that operations of conversion
(operations) are PPT-PMM.  Then we will maximize the following quantity:
\be
F = \tr[P_{00}^{\otimes m} \Lambda (\rho^{\otimes n})].
\label{fid1}
\ee
The fidelity $F$ is a function of $n$, since the rate $r$ is fixed.
 Our general argument will be the following. 
If for given rate  $F$ optimized over such operations is smaller than one, then the rate is not  
achievable.
The smallest such (achievable) rate is the upper bound for the optimal achievable rate, hence for $I_L$. 

Optimization of $F$ will have two stages:  first we will change the problem of optimization 
over $\Lambda$ to optimization over set of some positive operators $\Pi$, which fulfill some
(rather complicated) conditions. The optimization over those constraints 
is an example of so-called semidefinite program.  In second stage by 
duality method  used in semidefinite programming
we will find the bound on F expressed as infimum over Hermitian operators (without 
any additional constraints). We will then obtain bounds for localisable information 
for Werner and isotropic states by choosing appropriate Hermitian operator or by optimizing over a class of
Hermitian operators.   

It is useful to observe, that since $\Lambda$ is a CPTP map we have
\be
\tr [P_{00}^{\otimes m} \Lambda(\rho^{\otimes n})]=
\tr [P_{00}^{\otimes m} \sum_i V_i \rho^{\otimes n} V_i^{\dagger}] =
\tr [\sum_i V_i^{\dagger} P_{00}^{\otimes m} V_i \rho^{\otimes n} ], 
\ee
where $V_i$ are Kraus operators of the map.
We used here the fact $\tr AB = \tr BA$ for any operators A and B. The 
map $\sum_i V_i^{\dagger} (.) V_i \equiv  \Lambda^{\dagger}$
is called dual map (with respect to $\Lambda$).
It is clearly a CP map too (yet it need not be trace preserving).
The  meaning of dual maps to  NLOCC operations is exhibited in \cite{ustron}.
Here we are interested in the following operator
\be
\Pi=\Lambda^{\dagger}(P_{00}^{\otimes m}) .
\label{pi}
\ee
We can write fidelity by means of $\Pi$ as follows
\be
F=\sup_\Pi \tr [\rho^{\otimes n} \Pi]
\ee
where supremum is taken over all $\Pi$ of the form  (\ref{pi}).
Let us now prove the following fact, which amounts to first stage of optimization.
\begin{fact}
\label{fidelity-primal}
For given rate $r$ and the number of input copies n of any state $\rho \in {\cal C}^{d} 
\otimes {\cal C}^{d}$, the optimal fidelity is bounded by
\be
 F \leq \sup_{\Pi} \tr [\Pi (\rho^{\otimes n})],
\ee
where
\be
\nonumber
 0\leq \Pi \leq \id ,\hskip0.3cm \Pi^{\Gamma} \geq 0 ,\hskip0.3cm \tr [\Pi] = 2^{n(2\log  
d-r)}\equiv  K.  
\ee  
\end{fact}

{\bf Proof.} We need to show that $\Pi$ satisfies the displayed constraints.
To this end we will need some properties of partial transposition, 
which we write down here for clarity. Namely for any operators A and B one has 
\ben
&&(\Gamma_1) \quad \tr A^{\Gamma} = \tr A \\ \nonumber
&&(\Gamma_2) \quad \tr A^{\Gamma} B = \tr A B^{\Gamma}\\ \nonumber
&&(\Gamma_3) \quad \tr (A B)= \tr A^{\Gamma} B^{\Gamma} \\ \nonumber
&&(\Gamma_4) \quad (A^\Gamma)^\Gamma = A \\ \nonumber
&&(\Gamma_5) \quad  \Gamma \mbox{ preserves hermiticity}
\een
Now, let us first check if the $\Pi$ defined above operator fulfills the stated constraints.
Actually, we will see, that $0\leq \Pi \leq \id$ is a
consequence of the fact that $\Lambda$ is CPTP map and that $P_{00}^{\otimes m} \leq \id$, 
and positivity of  $\Pi^{\Gamma}$ is a consequence of $\Lambda$ being PPT map and the 
fact that ${(P_{00}^{\otimes m})}^\Gamma \geq 0$. We will use $P$ instead of $P_{00}^{\otimes m}$ 
for convenient notation.

Positivity of $\Pi$ is rather clear, since  - as it came up - $\Lambda^{\dagger}$ is a 
positive map. 
Comparing $\Pi$ with identity is simple, too. For any state $\sigma$ we have 
$\sigma \leq \id  $, which implies that for any positive operator A 
\be
\tr [\Lambda(A) \sigma] \leq \tr \Lambda(A) = \tr A,
\ee
where the equality expresses the fact that $\Lambda$ is trace preserving.
 This however is equivalent to the 
\be
\tr [A \,\Lambda^{\dagger}(\sigma)] \leq \tr[ A \,\id]
\ee
which for $\sigma = P$ gives  $\Pi \leq \id$.  
To check positivity of partially transposed $\Pi$ we need to show, that for any state $\sigma$
\be
\tr \left[\sigma [\Lambda^{\dagger}(P)]^{\Gamma}\right]\geq 0.
\ee
Applying $(\Gamma_2)$ one gets 
\be
\tr [\sigma^{\Gamma} \Lambda^{\dagger}(P)] \geq 0,
\ee
what by definition of dual map is equivalent to 
\be
\tr [ \Lambda( \sigma^{\Gamma})P] \geq 0.
\ee
Applying subsequently $(\Gamma_1)$ and $(\Gamma_3)$ one ends up with
\be
\tr [(\Gamma \Lambda \Gamma)(\sigma) P^{\Gamma}] \geq 0,
\ee
which is true, since both $(\Gamma \Lambda \Gamma)(\sigma)$ and $P^{\Gamma}$ are positive
operators. First because $\Lambda$ is a PPT map and second because $P$ is a product state
for which Peres separability criterion guarantees positivity of partial transposition. 

To prove the last property of $\Pi$ 
we use the fact, that $\Lambda$ is PMM i.e. $\Lambda({\id \over d_{in}}) = {\id \over d_{out}}$.
We then obtain,
\be 
\tr \Pi \equiv \tr\Lambda^{\dagger}(P)\id_{in} =  
d_{in} \tr P \Lambda({\id_{in} \over d_{in}}) = d_{in} \tr P {\id \over d_{out}} = 
{d_{in} \over d_{out}} = K,
\ee   
where $d_{in} = d^{2n}$ and $d_{out} = 2^{2m}$. This ends the proof. \blacksquare

 

Now, in second stage of optimization of the fidelity,
we can rearrange our task, using the concept of duality in semidefinite
programming. We will need the notion of positive part of 
operator. For Hermitian operator $H$  positive $H_+$ and negative $H_-$ parts are defined by 
 $H_+ - H_- = H$ and $H_+H_- = 0$, i.e. 
$H_+ = \sum_i \lambda_i^+ |\psi_i^+\> \<\psi_i^+|$, 
$H_- = \sum_i \lambda_i^- |\psi_i^-\> \<\psi_i^-|$ where $\lambda_i^+$ $(\lambda_i^-)$ are 
nonnegative (negative) eigenvalues and $\psi_i^\pm $ are corresponding eigenvectors.

{\theorem
For any state $\rho$ acting on $ ({\cal C}^{d} \otimes {\cal C}^{d})^{\ot n}$ 
and a  fixed rate $r$
\be
F \leq \inf_{D}[ \tr (\rho - D)_+ + K \lambda_{max}(D^{\Gamma})]
\ee
where $\lambda_{max}(D^{\Gamma})$ is the maximal eigenvalue of hermitian 
operator $D^\Gamma$; and  $K=2^{n(2\log d -r)}$.
}

{\it Proof.}
By just adding and subtracting proper terms which is similar to the 
Lagrange's multipliers method, using $(\Gamma_2)$ we can state that for any operators $A$, $B$   
and
for any real parameter $\lambda$, we have
\be
	\tr \Pi\rho = \tr A - \tr (-\rho + A - B^{\Gamma} + \lambda \id )\Pi + \lambda K
-\tr A(\id -\Pi) - \tr B\Pi^{\Gamma} + \lambda(\tr \Pi - K)
\ee

Now if A and B are positive operators and $ A \geq \rho + B^{\Gamma} - \lambda \id  $ we have:
\be
\tr \Pi\rho \leq \tr A + \lambda K,
\ee
since absent terms in LHS are non-positive according to the constraints on $\Pi $. Then
\be
\sup_{\Pi} \tr \Pi\rho \leq \inf_{A,B, \lambda}\tr [A + \lambda {K \over d^2} \id ]
\ee
where $A \geq 0,\hskip0.2cm B \geq 0,\hskip0.2cm A-B^{\Gamma} + \lambda \id  \geq \rho, 
\lambda \in R$. By introducing new variable $D = \lambda \id  - B$ it can be changed 
into the following form:
\be
F \leq \inf_{{{A \geq 0, B \geq 0\atop A\geq \rho - D^{\Gamma}} \atop D = \lambda \id -B, 
\lambda \in R}} \tr [A + \lambda {K \over d^2} \id ].
\ee
Taking subsequently infimum over $D$, $A$ and $B$ we obtain
\be
F \leq \inf_{D} {\large \{}
\inf_{{A \geq 0 \atop A\geq \rho - D^{\Gamma}}}[\tr A] \hskip0.2cm +\hskip0.2cm 
\inf_{{B \geq 0 \atop \lambda \id  = D +B, \lambda \in R}}[\lambda {K \over d^2} \id ]
{\large \}},
\ee
where D (as a combination of positive operators) is a Hermitian operator.
Having D fixed one can easily minimize two separate terms over A and B respectively.
Concerning the first term, since $A \geq 0$ and $A \geq \rho - D^{\Gamma}$, the eigenvalues 
$\lambda^{A}_i$ of A must be greater than zero, and greater than the eigenvalues
$\lambda^{\rho, D}_i$ of  
the  $\rho - D^{\Gamma}$ operator. Thus we have $\lambda^{A}_i=\max(\lambda^{\rho, D}_i,0)$
which gives:

\be
\inf_{{A \geq 0 \atop A\geq \rho - D^{\Gamma}}}[\tr A] =\tr  (\rho - D^{\Gamma})_+
\ee

Turning now to the second term, one can see, that $\lambda \id  - D \geq 0$, hence $\lambda$ must  
be not less
than maximal eigenvalue of D, thus we end up with:

\be 
\inf_{{{B \geq 0 \atop \lambda \id  = D +B}\atop \lambda \in R}}[\lambda {K \over d^2} \id ]=
K \lambda_{max}(D).
\ee
This leds to the formula
\be
F \leq \inf_D{\{\tr  (\rho - D^{\Gamma})_+ + K \lambda_{max}(D)\}}.
\label{zmiana_D}
\ee
By $\Gamma_4$ and $\Gamma_5$ any Hermitian operator $D$ is of the form $\tilde{D}^\Gamma$,
where $\tilde{D}$ is also Hermitian operator, hence we can rewrite the formula in the 
following way
\be
F \leq \inf_D{\{\tr  (\rho - D)_+ + K \lambda_{max}(D^{\Gamma})\}},
\ee 
which ends the proof. \blacksquare

The above result, gives us the condition on F with much simpler constraints - 
we optimize over set of hermitian operators. Analogously, one can   prove 
similar theorem for $\rho \in ({\cal C}^{d1} \otimes {\cal C}^{d2})^{\ot n}$.


\section{Bound for rate of concentrating information}
\label{sec:bound-rates}
In previous section we showed that the  fidelity of  concentrating information by NLOCC is bounded  
by 
\be
F\leq \inf_{D} Tr(\varrho^{\otimes n} - D)_{+} + 2^{n(2log_{2}d-r)} \lambda_{max}(D^{\Gamma})
\ee
Starting from this inequality we can find two bounds for  the rate $r$ (by PPT-PMM operations) denoted by  $B_1$
 and $B_2$. The bound $B_1$ is weaker than $B_2$, however we derive it saperately, as the proof is more transparent than that for $B_2$.  
\begin{theorem}
\label{tw1}  
For any states $\varrho$ 
\be
r\leq 2\log_{2}d+\log_{2}\lambda_{max}(|\varrho^{\Gamma}|)\equiv B_{1}(\varrho)
\label{nier}
\ee
\end{theorem}
$\mathnormal{Proof}$. 
We will show that inequality  (\ref{nier}) must be true to make the fidelity converge to 1.
Let us take $D=\varrho^{\otimes n}$.
Then
\be 
F\leq \limsup_{n\rightarrow \infty} 2^{n(2\log_{2}d-r)} \lambda_{max}((\varrho^{\otimes  
n})^{\Gamma})
\label{F}
\ee
The requirement $F\rightarrow 1$ is equivalent to condition
\be
\limsup_{n\rightarrow \infty} n(2\log_{2}d-r)+ \log_{2}\lambda_{max}((\varrho^{\otimes  
n})^{\Gamma})\rightarrow 0
\ee
It implies
\be
r\leq 2\log_{2}d+\limsup_{n\rightarrow  
\infty}\frac{1}{n}\log_{2}\lambda_{max}(\varrho^{\Gamma})^{\otimes n}
\ee
Notice that 
\beng
\limsup_{n\rightarrow \infty}\frac{1}{n}\log_{2}\lambda_{max}(\varrho^{\Gamma})^{\otimes n}&=&
\limsup_{n\rightarrow \infty}\frac{1}{n}\log_{2}(max|\lambda(\varrho^{\Gamma})|)^{n}\\
=\limsup_{n\rightarrow \infty}\log_{2}\lambda_{max}(\varrho^{\Gamma})&=&
\log_{2}\lambda_{max}(\varrho^{\Gamma})
\eeng
Then we have 
\be
r\leq 2\log_{2}d+\log_{2}\lambda_{max}(|\varrho^{\Gamma}|)
\ee
This ends the proof.

\begin{theorem}
\label{tw2}  
For any states $\varrho$ and $\sigma$,
\be
r(\varrho)\leq 2\log_{2}d+S(\varrho|\sigma)+\log_{2} \lambda_{max}(|\sigma ^{\Gamma}|)
\equiv  
B_{2}(\varrho,\sigma)
\ee
\end{theorem}
${\bf Remark }$ 1. Notice, that $\lambda_{max}(|\varrho^{\Gamma}|)
=||\sigma^{\Gamma}||_{op}$, where $||A||_{op}$ is operator norm. Then the bound can be written as:
\be
B_{2}(\varrho,\sigma)= 2\log_{2}d+S(\varrho|\sigma)+\log_{2} ||\sigma^{\Gamma}||_{op}
\ee
It is interesting to compare this expression this formula  with Rains bound for 
PPT distillable entanglement D :
\be
D \leq S(\varrho|\sigma)+\log_{2}||\sigma^{\Gamma}||_{Tr}
\ee
${\bf Proof}$. 
Let $S=S(\varrho|\sigma)$ and  
$L=\log_{2}\lambda_{max}(|\sigma^{\Gamma}|)=\log_{2}||\sigma^{\Gamma}||_{op}$. We will show that  
if 
\be
r-2\log_{2}d \equiv  x > S+L
\ee
 then F cannot converge to 1.
We have
\be
F \leq Tr(\varrho^{\otimes n} - D)_{+} + 2^{-nx} \lambda_{max}(D^{\Gamma})
\ee
Let us take 
\be
D=2^{ny}\sigma^{\otimes n}
\ee
where $S < y < x- L$ . (We can find such y, because $x > S+L$)\\
Notice that 
\be
2^{-nx}\lambda_{max}[(2^{ny}\sigma^{\otimes n})^{\Gamma}]\leq 2^{n(y-x+L)}
\ee
Then 
\be
F \leq Tr(\varrho^{\otimes n} -2^{ny}\sigma^{\otimes n} )_{+} + 2^{n(y-x+L)}
\label{eq-fidelity}
\ee
$2^{n(y-x+L)}$ converges to 0 
because 
\be
y-x+L < 0
\ee
The first term in (\ref{eq-fidelity})
cannot converge to 1 because $y >S(\varrho|\sigma)$, as shown by Rains \cite{Rains2001} .
 (It follows from quantum Stein lemma, see e.g. \cite{Stein}.)
This ends the proof. 

$\bf{Remark}$ 2
To obtain the full strength of bound of Theorem \ref{tw2}, one should optimize the choice of  
$\sigma$.
In what follows, we will say that $\sigma$ is optimal for $\varrho$ if 
\be
B_{2}(\varrho,\sigma)=\min_{\sigma'}B_{2}(\varrho,\sigma')\mathop{=}\limits^{\textrm{df}}
B_{2}(\varrho)
\ee
where $\sigma'$ ranges over all states.
 
\section{Results for Werner and isotropic state.}

In this section we will find bounds for rate for states possesing high symmetry: Werner states and isotropic ones. We will compare bounds $B_1$ and 
 $B_2$ with one other. 


Let us  start with Werner state. We describe our results for  Werner state of the form (\ref{wer1}).
Using  Theorem \ref{tw1} we obtained the following bound:
\ben  
B_{1}= 
\left \{
\begin{array}{ll}
2\log_{2}d-\log_{2}(d^{2}+d\beta) & \textrm{for}  -\frac{2}{d} < \beta < 0 \\ 
2\log_{2}d+\log_{2}|\frac{1+d\beta}{d^{2}+d\beta}| & \textrm{for}  \quad  0 \leq \beta \leq 1 
\quad and \quad  -1 \leq \beta \leq -\frac{2}{d}
\end{array}
\right. 
\een

If we want to find the bound using  Theorem \ref{tw2} we have to  optimize  
$B_{2}(\varrho,\sigma)$.
 Luckily, as in \cite{Rains2001}, it boils down to minimizing only over  Werner states. It is due  to the following 
two facts. First, any state $\varrho$
 if subjected to random transformation of form $U\otimes U$(called $U\otimes U$ twirling) becomes  
Werner state.
 \be
 \int  U \otimes U \varrho U^{\dagger}\otimes U^{\dagger}dU = \varrho_{W}
 \ee
 Second, value of $B_{2}(\varrho,\sigma)$ is nonincreasing after twirling operation.
Third, $B_{2}(\varrho,\sigma)$ is convex function,  
because the quantities 
$S$ and $L$ possess these properties.
 Then for any state $\sigma$
\ben  \nonumber
&&B_{2}(\varrho_{W},\sigma) = \int B_{2}(\varrho_{W}, \sigma)dU=
\int B_{2}(U \otimes U \varrho_{W} U^{\dagger}\otimes U^{\dagger},U \otimes U \sigma  
U^{\dagger}\otimes U^{\dagger})dU \geq \\ \nonumber
&& B_{2}(\int U \otimes U \varrho_{W} U^{\dagger}\otimes U^{\dagger}dU,\int U\otimes U \sigma  
U^{\dagger}\otimes U^{\dagger}dU)
= B_{2}( \varrho_{W} , \int U\otimes U \sigma U^{\dagger}\otimes U
^{\dagger}dU )= B_{2}( \varrho_{W} , \sigma_{W} ) 
\een
where $\sigma_{W}$ is a Werner state.
Thus, we can see that for any state $\sigma$ we can find such Werner state $\sigma_W$, which gives  
no grater value of $B_2$ than $\sigma$.
This fact simplifies our calculation to optimize $B_{2}(\varrho_{W},\sigma)$ on Werner state. Now,  
we can find the smallest value of 
$ B_{2}( \varrho_{W} , \sigma_{W} )$, where $\varrho_{W}$ is given by the formula (\ref{wer1}) and  
$\sigma_{W}$ is of the following form:
\be
\varrho_{W}=\frac{1}{d^{2}+d\alpha}(I+\alpha V),\quad -1\leq \alpha \leq 1
\ee
In this case $B_{2}(\varrho_{W},\sigma)$ is a function of three parameters: $d$, $\beta$ and  
$\alpha$, 
where the first two parameters are fixed. So to optimize $B_{2}(\varrho_{W},\sigma)[\alpha]$ it is  
enough 
to find an minimum of this function depending on $\alpha$.
This way  we obtain the following value of  
$B_{2}(\varrho)$:
\ben  
B_{2}=\left \{
\begin{array}{ll}
2\log_{2}d-S(\varrho_{W})-\frac{d^{2}+d}{2}\frac{1+\beta}{d^{2}+d\beta}\log_{2}\frac{d-2}{d}
-\frac{d^{2}-d}{2}\frac{1-\beta}{d^{2}+d\beta}\log_{2}\frac{d+2}{d}
& \textrm{for}  -1\leq \beta < \frac{-3d}{d^2+2} \\ 
2\log_{2}d-S(\varrho_{W})-\frac{d^{2}+d}{2}\frac{1+\beta}{d^{2}+d\beta}\log_{2}{(1+\alpha)}
-\frac{d^{2}-d}{2}\frac{1-\beta}{d^{2}+d\beta}\log_{2}{(1-\alpha)}
& \textrm{for} \frac{-3d}{d^2+2}\leq \beta <  \frac{-1}{d}\\ 
2\log_{2}d -S(\varrho_{W}) & \textrm{for} \quad \frac{-1}{d}\leq \beta \leq 1  
\label{b2w}
\end{array}
\right. 
\een
$\textrm{where}\quad  \alpha =\frac{1+d\beta}{d+\beta}.$ The entropy  $S(\varrho_{W})$  is given by:
\ben
S(\varrho_{W}) =-\frac{d^{2}-d}{2}\frac{1-\beta}{d^{2}+d\beta}\log_{2}\frac{1-\beta}{d^{2}+d\beta}-\frac{d^{2}+d}{2}\frac{1+\beta}{d^{2}+d\beta}\log_{2}
f\frac{1+\beta}{d^{2}+d\beta}
\een
We have obtained two upper bounds for amount of information, we can localize. 
Of course $B_2$ is always not worse than $B_1$, so we  will consider $B_1$ only to compare with $B_2$. 
Now, we would like to find  a lower bound for $I_l$. 
Consider  some  NLOCC (one-way ) protocol $\mathcal{P}$ for concentrating information to local  
form.
The amount  of information  we can concentrate using  $\mathcal{P}$ is a lower bound for $I_l$.
Our  protocol $\mathcal{P}$ is following:
(i) Alice makes an optimal complete von Neumann measurement represented by $P_i=|i\>\<i|$ on her  
subsystem.
(ii) After that  she sends her part to Bob. Alice can do this, because  after   measurement her  
part of  state  is classical-like and 
classical channel does not destroy it, if we do it adequately, i.e. sometimes before sending we perform 
some unitary operation to avoid changing the state by the channel.
(iii) Bob  upon receiving the whole state can extract $2\log_2d-S(\varrho'_{AB})$ bits of information, where  
$\varrho'_{AB}$ is obtained from 
$\varrho_{AB}$ by  Alice's operation (i), (ii).

\begin{figure}
\centerline{\psfig{file=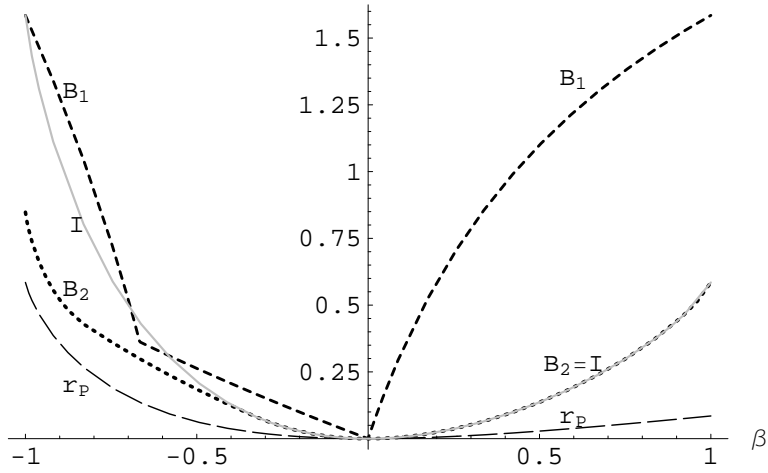}}
\caption{The dashed lines represent bounds of rate for Werner states (d=3) ($r_{\mathcal{P}}^{\rightarrow} \leq 
B_2 \leq  B_1$) and continuous line represents quantity described by $I=2\log_2d-S(\varrho)$. Note that $B_2$ is equal to 
information content of state in whole range of separability ($\beta \geq -\frac{1}{3}$) \label{rys1}}
\end{figure}

\begin{lemma}
For $d \otimes d$ state  $\varrho_{AB}$ with  maximally mixed subsystem A by
use of the protocol $\mathcal{P}$, we can concentrate to local form  
$r_{\mathcal{P}}^{\rightarrow}$ information, where $r_{\mathcal{P}}^{\rightarrow}$
is described by:
\ben 
&&r_{\mathcal{P}}^{\rightarrow}= \sup_{P_{i}}(\log_{2}d - \sum_{i}p_{i}S(\varrho'^{i}_{B})) \\  
\nonumber
&&\textrm{where}\quad p_{i}=tr (\varrho_{AB}P_{i}\otimes I ) 
\quad \textrm{and}\quad \varrho_{B}^{i}=\frac{1}{p_i} tr_A (P_i\otimes I \varrho_{AB} P_{i}\otimes  
I) 
\een 
\end{lemma}
$\textbf{Proof.}$ After sending by Alice her part, Bob  possesses the  whole state and can extract  
$2\log_2d-S(\varrho'_{AB}) $, where
$\varrho'_{AB}=\sum_{i}p_i  
|i\>\<i|\otimes \varrho^{i}_{B}$. The states $|i\>$ are orthogonal  and it implies that 
$S(\varrho'_{AB})=H({p_i})+ \sum_{i}p_{i}S(\varrho'^{i}_{B})$. Shannon entropy $H({p_i})$ is  
amount to entropy of Alice's part 
after her  measurement. We know, that entropy cannot decrease after measurement but also cannot  
increase, because is maximal.
It implies  that $H({p_i})=\log_{2}d$. Then we have 
\ben \nonumber
r^{\rightarrow}_{\mathcal{P}}&=& 2\log_2d-S(\varrho'_{AB})
=2\log_{2}d-(\log_{2}d+\sup_{P_{i}} (\sum_{i}p_{i}S(\varrho'^{i}_{B})))=\\ \nonumber
&=&\log_{2}d -\sup_{P_{i}}(\sum_{i}p_{i}S(\varrho'^{i}_{B}))
\een
This ends the proof.\\
For Werner states $r^{\rightarrow}_{\mathcal{P}}$ is achieved by any measurement of Alice. 
It follows from the fact that $\varrho_W$ is $U\otimes U$ invariant.
We obtain:
\ben
r^{\rightarrow}_{\mathcal{P}}(\varrho_W)= \log_{2}d +  
\frac{1+\beta}{d+\beta}\log_{2}(1+\beta)-\log_{2}(d+\beta)
\een
Let us here compare the bounds for amount of localizable information with each other.\\
Fig. \ref{rys1} shows lower and upper bounds for rate in comparison to information content of  
state.
 For Werner states bound $B_{2}$ is much better than $B_{1}$. 
For separable state $B_{2}$ is trivial, it coincides  with information contents of state
$I=2\log_{2}d - S(\varrho)$. For entangled states it is better than
I.
%


\begin{figure}[h]
\centering
\centerline{\psfig{file=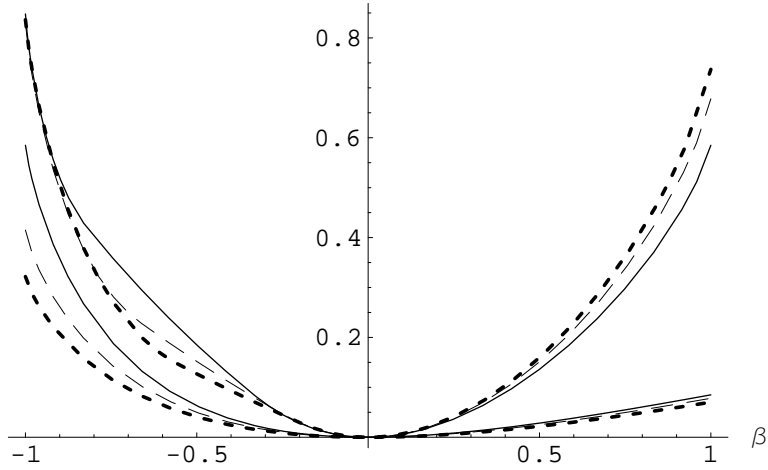}}
\caption{Upper bound $B_2$ and lower bound  $r^{\rightarrow}_{\mathcal{P}}$ of rate for Werner states  
(d=3,4,5). 
\label{rys2}}
\end{figure}

Looking at Fig. \ref{rys2} we can see $B_2$ and $r^{\rightarrow}_{\mathcal{P}}$ for some  
different dimensions of Hilbert space of 
Werner state. Continuous lines represent bounds for d=3 , the long dashed lines bounds for d=4 
and the short dashed for d=5.

Now, let us present results for isotropic state. For these  states the bound $B_1$ is given by:
\ben 
B_{1}= \left \{
\begin{array}{ll}
\log_{2}[-\lambda(d+1)+1] & \textrm{for} \quad \lambda<0 \\
\log_{2}[\lambda(d-1)+1] & \textrm{for} \quad \lambda\geq 0
\end{array}
\right.
\een
Using the same arguments like for Werner state we can show that if we want to find value of
$B_2(\varrho_{iso})$ we ought to optimize $B_{2}(\varrho_{iso},\sigma)$ on isotropic state. 
Analogously, as  in previous case, we can find out that :
\ben 
B_{2}=  \left \{
\begin{array}{ll}
2\log_{2}d-S(\varrho_{iso})&
 \textrm{for} \quad \frac{-1}{d^{2}-1} \leq \lambda \leq \frac{1}{d+1}\\ 
2\log_{2}d-S(\varrho_{iso}) +\log_{2}\frac{1+p(d-1)}{1-p}
-\frac{1+\lambda(d^{2}-1)}{d^{2}}\log_{2}\frac{1-p}{1+p(d^{2}-1)} &
\textrm{for} \quad \frac{1}{d+1}\leq \lambda \leq 1 \\  
\end{array}
\right.
\een
$\textrm{where}\quad p=\frac{(d+1)\lambda -1}{(1-d^{2})(\lambda-1)+d}$. The entropy of isotropic state is given by:
\ben
S(\varrho_{iso})=-\frac{1+\lambda(d^{2}-1)}{d^{2}}\log_{2}\frac{1+\lambda(d^{2}-1)}{d^2}-
\frac{(1-\lambda)(d^{2}-1)}{d^2}\log_{2}\frac{1-\lambda}{d^2}
\een
Isotropic state possess similar properties like Werner states, so if we want 
to obtain a  value  of $r_{\mathcal{P}}^{\rightarrow}$ we should proceed similarly 
like for that family of state. Then we have
                                                                                                                                                                                                                                         
\begin{figure}
\centerline{\psfig{file=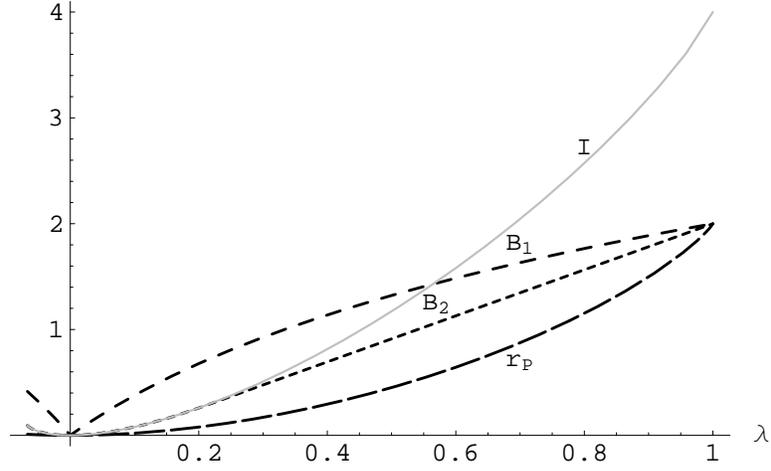}}
\caption{The dashed lines represent bounds of rate for isotropic states (d=3) and continuous line  
information content of state. Note that $B_2$ is equal to $I=2\log_2d-S(\varrho)$ for
 whole range of separability ($\lambda \leq \frac{1}{1+d}$) \label{rys3}}. 
\end{figure}

\ben
r_{\mathcal{P}}^{\rightarrow} = 
\log_{2}d+(\lambda+\frac{1-\lambda}{d})\log_{2}(1+\frac{1-\lambda}{d}) 
\een  
For isotropic state the bound $B_{2}$ is  better again than $B_{1}$ and also  only for entangled  
isotropic state the upper bound is nontrivial.
The upper and lower bounds agree for $P_{+}$ and are obviously equal to $\log_{2}d$. The bounds and information content are compared on  figure  
(\ref{rys3}) for d=3. 
The upper dashed line represent $B_2$, the lower $\mathop{r}_{\mathcal{P}}^{\rightarrow}$. The gray  
continuous line is a information content of state, i.e. $2\log_2d-S(\varrho_{iso)}$.

\begin{figure}
\centerline{\psfig{file=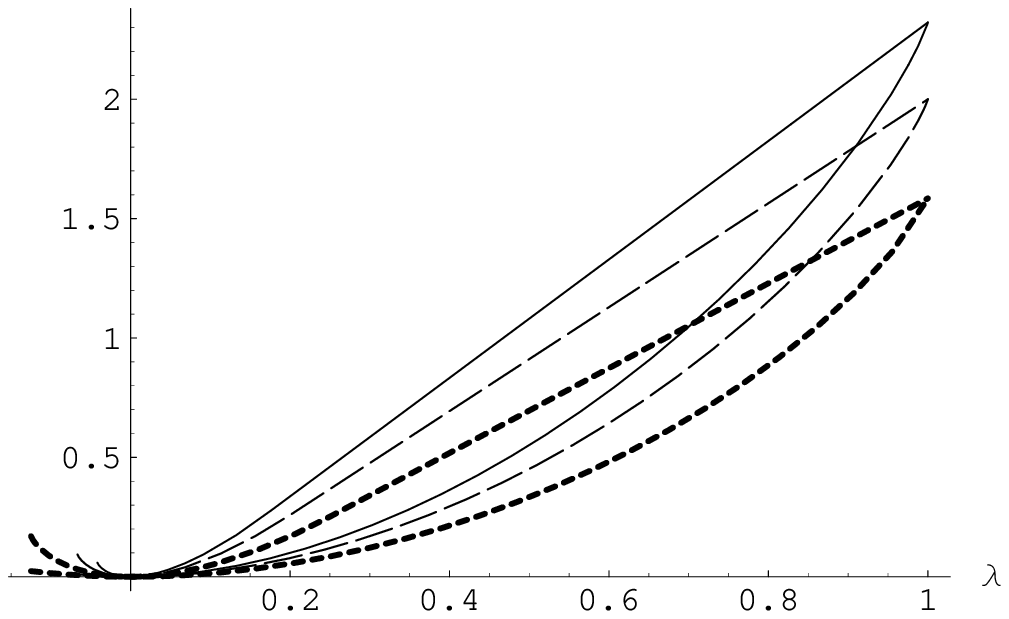}}
\caption{Upper bound B2 and lower bound  $r^{\rightarrow}_{\mathcal{P}}$ of rate for isotropic  
states (d=3,4,5). 
\label{rys4}}
\end{figure}

On Figure  \ref{rys4}
 give us ability to compare quantity B2 and   
$\mathop{r}_{\mathcal{P}}^{\rightarrow}$ for some different dimension 
(d=3,4,5).

\section{Comparison quantum deficit with  measures of entanglement}

As one  knows that quantum deficit $\Delta$ can be treated as  a measure of quantum correlations \cite{OHHH2001}.
Having bound  localisable   information we can find bounds for $\Delta$. We can do this, because  
quantum deficit is defined as a difference  between total information and information $I_l$, which  
can be localized by NLOCC (\ref{delta}) and we know that $I_l$ is bounded by information localisable by  
using PPT-PMM operation. 

We  would like to compare quantum deficit with another measure of entanglement, because we  
suppose that this quantity 
is more general measure of "quantumness" of state that well-known measure of entanglement. 

This way we get lower bound  $\delta_B$  for $\Delta$:
\ben 
\delta_B(\varrho)=2\log_2d-S(\varrho)-B_2 
\een
and upper bound $\delta_P$   :
\ben 
\delta_P  (\varrho)=2\log_2d-S(\varrho)-r^{\rightarrow}_{\mathcal{P}}
\een
We can compare  $\Delta$ with  known measures of entanglement.\\ 
The regularized relative entropy of entanglement $E_R^\infty$ for Werner states is
given by \cite{werre}:
\ben
E_R^{\infty} =
\left \{
\begin{array}{ll}
\log_2 \frac{d-2}{d}+\frac{(d-1)(1-\beta)}{2(d+\beta)}\log_2\frac{d+2}{d-2} & \textrm{for} \quad  
-1\leq \beta \leq -\frac{3d}{d^2+2} \\
1+H(\frac{(d-1)(1-\beta)}{2(d+\beta)}) & \textrm{for} \quad -\frac{3d}{d^2+2}\leq \beta \leq   
-\frac{1}{d} \\
0 & \textrm{for} \quad -\frac{1}{d}\leq \beta \leq 1    
\end{array}
\right. 
\label{dest}
\een
Entanglement of formation is described by the following formula:
\ben
E_F=
\left \{
\begin{array}{ll}
H(\frac{1}{2}(1-\sqrt{1-(\frac{1+d\beta}{d+\beta})^2} \big{)} & \textrm{for} \quad -1\leq \beta  
\leq -\frac{1}{d} \\
0 & \textrm{for} \quad -\frac{1}{d} < \beta \leq 1 
\end{array}
\right. 
\label{ef}
\een

\begin{figure}
\centerline{\psfig{file=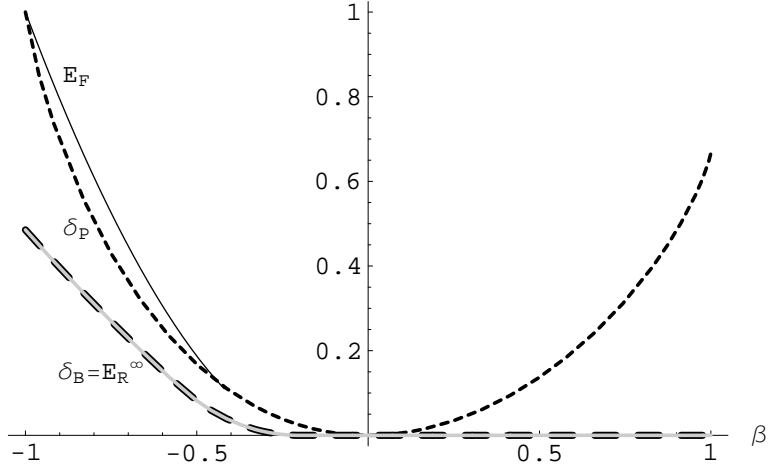}}
\caption{The dashed lines represent upper and lower bounds of $\Delta $ for Werner states (d=5),
grey continuous regularized relative entropy of entanglement  $E_R^\infty $ and black continuous entanglement  
of formation $E_F$.  \label{rys7}} 
\end{figure}

We can see on Fig. (\ref{rys7}) the graphs of $\delta_B$  , $\delta_P$, $E_F$ and $E_R^\infty$ for Werner states.  
We obtain that 
$\delta_B$   and $E_R^\infty$ are equal. For Werner states with $\beta < 0.42$ we  
have that  quantum deficit  is not less  than entanglement of formation:  $\Delta\leq E_F$.

\begin{figure}[h]
\centerline{\psfig{file=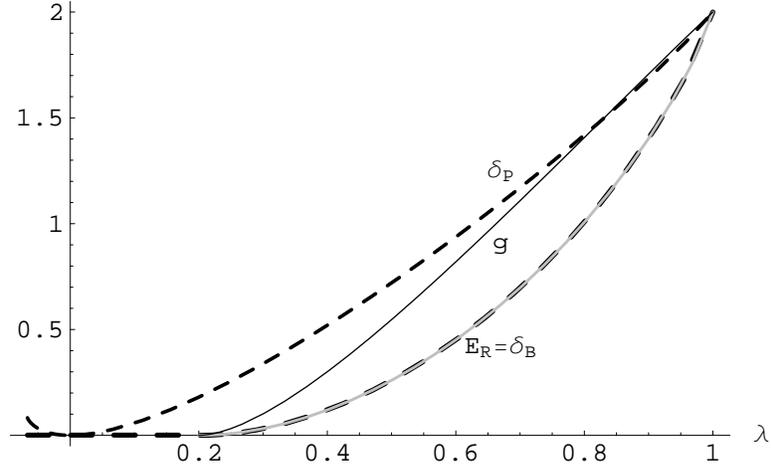}}
\caption{The dashed lines represent bounds for  $\Delta $  for isotropic states (d=3), 
grey continuous regularized relative entropy of entanglement $E_R$ and black continuous  $g(\gamma )$(whose convex hull is
entanglement of formation $E_F$).
\label{rys8}} 
\end{figure}

Let us now pass to isotropic states. For entangled  ones with parameter 
$\lambda \in (\frac{1}{d+1},1)$ the relative entropy of entanglement $E_R$   is given by: 
\ben
E_R=\log_2 d  + f \log_2 f + (1 - f) \log_2 \frac{1 - f}{d - 1} 
\label{ef}
\een
where $f=\frac{(d^2 - 1)\lambda  + 1}{d^2}$. For another isotropic states it is surely zero. The formula of   
entanglement of formation we can find in paper \cite{VW01}.
For nonseparable states it is in form of:
\ben
E_F(\varrho)&=& co(g(\gamma )) \\
g(\gamma )&=&(H_2(\gamma)+(1-\gamma)\log_2(d-1))
\een
where  
$\gamma=\frac{1}{d^2}\Big{(}\sqrt{d\lambda+\frac{1-\lambda}{d}}+\sqrt{(d-1)\frac{(d^2-1)(1-\lambda
)}{d}}\Big{)}^2$ and 
{\bf co} means a convex hull \cite{VW01}. On figure (\ref{rys8}) we can see graphs of two measure  
of entanglement and bounds for delta.

We can notice that the graphs of $\delta_B$  agree with $E_R$. Similary as for Werner states we have $\delta_B=E_R$. $\delta_P$ is grater than 
$E_F$ for most isotropic states. (We do not know, if it is true for all isotropic states).
For maximally entangled  state $P_+$ all these quantities are equal.
  
\section{Fidelity for distillation of local pure states and singlets}

\begin{figure}
\centerline{\psfig{file=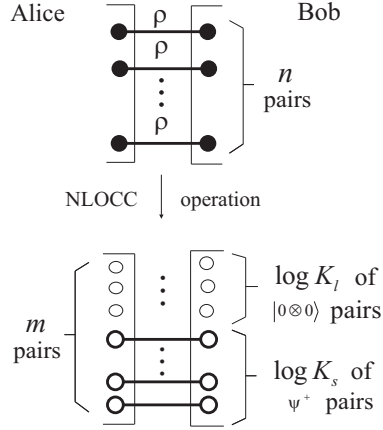,width=5cm}}
\caption{Scenario of distillation of pure states in their two extremal forms: purely local 
(product) states and purely nonlocal (maximally entangled) states. 
The fixed rates of transition gives the proportion of the number of input copies n 
of state $\rho$ to the numbers of output copies of {\it local } states ($\log K_l$) and 
singlet states ($\log K_s$).}
\label{fig1}
\end{figure}

The well known counterpart of a qubit which represents the unit of local information is 
ebit - one bit of entanglement, represented by singled state i.e. unit of nonlocal information.
It has been stated in \cite{compl} that these two forms of information are complementary. 
If one distills maximal possible amount of one type of information, possibility of gaining 
the second type disappears. The optimal protocol in case of pure initial state 
$\psi_{AB}^{\otimes n}$, in which both types are obtained with some
ratios, has been also shown there. We will find a  bound for the fidelity
of such transition, in which both qubits and ebits are drawn in an NLOCC protocol in case
of general mixed state $\rho_{AB}^{\otimes n}$. One can
view this as a {\it purity distillation protocol}, because purity in general has two
extreme forms: purely local, and purely nonlocal one. This is due to the fact, that any 
nonproduct pure state, is asymptotically equivalent to the singlet state under the set of 
NLOCC operations \cite{nlocc}.
To this end - as before - we will consider broader class than the NLOCC, 
namely the class PPT-PMM. 
This time we have to fix two rates: the one which tell us how many pure local qubits we
would like to obtain ($r_l$), but also how many singlet states ($r_s$) will be acheived
per n copies of input state (see fig. \ref{fig1}).

Namely
\be
r_s = {\log{K_s}\over n},\hskip0.2cm r_l={2m - 2\log{K_s} \over n} 
\ee 
where $n$ is the number of input states, $m$ is the number of qubits which output states
occupies (both pure qubits, and singlets together) and $\log{K_s}$ is the output number of
singlet states. We get less than $2m \over n$ pure qubits since  $\log{K_s}$ singlets
use up $2\log{K_s}$ qubits out of $2m$ final. We should maximize the fidelity of transition:
\be
F = \tr[P_+^{\otimes \log K_s} \otimes P_{00}^{\otimes (m-\log {K_s})} \Lambda (\rho^{\otimes  
n})].
\ee
where $P_+^{\otimes \log K_s}$ is the projector onto the singlet state on 
${\cal C}^{K_s} \otimes {\cal C}^{K_s}$. Instead of tensor product of singlets and product 
states, we can equally well consider the output state 
as the same singlet state embedded in the larger Hilbert space ${\cal C}^{m} \otimes {\cal  
C}^{m}$.
Thus we shall maximize the following quantity:
\be
F = \tr[P_+^{K_s} \Lambda (\rho^{\otimes n})].
\ee
where $K_s$ reminds that $P_+$ is of less dimension than the whole Hilbert space it is 
embedded in. The rest of the space is occupied by pure local qubits i.e. the 
second resource drawn in this process.

Consequently we will first consider the fidelity in terms of the 
$\Pi = \Lambda^{\dagger}(P^{K_s}_+)$ operator where $\Lambda^{\dagger}$ is dual (hence CP)
to the $\Lambda$ which is CPTP from assumption.

Analogously as in section \ref{sec:fidelity-loc} we can obtain the following fact

\begin{fact}
For given rates $r_l$, $r_s$ and the number of input copies n, the optimal fidelity is given by
\be
 F \leq \sup_{\Pi} \tr [\Pi (\rho^{\otimes n})],
\ee
where
\be
\nonumber
 0\leq \Pi \leq \id ,\hskip0.3cm {-\id \over K_s} \leq \Pi^{\Gamma} \leq {\id \over K_s} 
,\hskip0.3cm \tr [\Pi] = 2^{n(2\log d - r_l - 2 r_s)}\equiv  K.  
\ee  
\end{fact}

Passing to dual problem, after some algebra we get 

{\theorem 
For any state $\rho$ acting on $({\cal C}^{d} \otimes {\cal C}^{d})^{\ot n}$ and 
rates $r_l$ and $r_s$ we have 
\be
F \leq \inf_{D}\{ Tr(
\rho - D)_+ + 
\inf_{\lambda} [{1 \over K_s} \tr |D^{\Gamma}-\lambda \id | + \lambda K ]\}
\ee
where infimum are taken over all hermitian operators D and all real numbers $\lambda$ 
respectively; $K=2^{n(2\log d - r_l - 2 r_s)}$ and $K_s=2^{nr_s}$.
}

It is not easy to obtain nontrivial results.
Suppose that we want to apply analogous ideas to those applied in section (\ref{sec:bound-rates}).  
Let us rewrite fidelity as follows (for some fixed $D$,
\be
F \leq Tr(\rho^{\ot n} - D) + 
2^{-nr_s} \tr |D^{\Gamma}-\lambda \id | + \lambda 2^{n(2\log d - r_l)}
\ee
The simplest approach would be to force  first  terms to vanish , and to try to put  
$D=\varrho^{\ot n}$. 
However, even by this simplification it is very difficult to find any bound for rates.

Yet, there are also "higher level" problems. Namely the main problem
within the connection between distillation of entanglement and the paradigm
of localising information, is whether distillation process consume
local information \cite{phase}.
It seems that our class of operations cannot feel this problem at all.
Indeed, it is likely, that any distillation process is a map
that preserve maximally mixed state \cite{jono}. Thus one should perhaps improve the approach,
by imposing more stringent constraints on class of operations.
This is because for initial maximally mixed state, we impose only
final maximally mixed state. However generically, final
dimension is smaller than initial one. This means that some
tracing out must take place, and we do not require the state
that was traced out must be maximally mixed. Thus in our class, pure
ancillas can be
added, under the condition that they are finally traced out.

\section{Discussion} 

 In this paper we have investigated {\it localisable information} and
associated {\it information
deficit} of quantum bipartite states. We used the fact, that localisable
information
can be defined as the amount of pure local qubits (per input copy)
that can be distilled by use of classical communication and local operations
that do not allow adding local ancilla in non-maximally mixed state.
We considered a larger class of operations which we called PPT-PMM
operations. They are
those PPT operations which preserve maximally mixed state. Then we managed
to formulate the problem of
distillation of pure product qubits in terms of semidefinite program.
Using duality concept in semidefinite programming we have found bound for
fidelity of transition given state
into pure product ones by PPT-PMM operations. In this way we obtained a
general upper bound
for amount of localisable information of arbitrary state. The bound was
denoted $B_2$
(we also obtained a simpler, but weaker bound $B_1$). It gives bound
$\delta_B$ for information
deficit. We were able to evaluate exactly the value of the bound $B_2$
for states exhibiting high symmetry - Werner states and isotropic states.
Quite surprisingly, the obtained related lower bound $\delta_B$ for
information deficit
turned out to coincide with relative entropy of entanglement in the case of
isotropic states, and with regularized relative entropy of entanglement
for Werner states. In other words: in those two cases, our bound for
information
deficit, turned out to be equal to Rains bound for distillable entanglement.
We also analysed a simple lower bound $r_P$ for localisable information,
and a parallel upper bound $\delta_P$ for information deficit.
We compared the latter bound with entanglement of formation.
In particular we obtained that for Werner states $(d=3)$, in entangled
region it is
strictly smaller than entanglement of formation. If one believes that
information
deficit is a measure of total quantumness of correlations, the conclusion
would be that
$E_F$ does not describe the entanglement present in state. Rather it includes
also the
entanglement that got dissipated during formation of the state.
Finally, we also discussed possibility of application of our approach to
the problem of simultaneous distillation of singlets and pure local states.
We provided bound for fidelity in this case.
However it is likely, that the chosen class of operations is too large
to describe information consumption in the process of distillation of
entanglement.
We believe that our results will stimulate further research towards
evaluating localisable information and information deficit. An important
open question is also the connection between information deficit
and entanglement measures. In particular, it is intriguing, how general is
the equality of our lower bound for deficit, and $E_r$ - upper bound for
distillable entanglement.

{\bf Acknowledgments}:
We thank Pawe\l{}  Horodecki,  Ryszard Horodecki and Jonathan Oppenheim for helpful discussion.
This work is supported by EU grants RESQ, Contract No. IST-2001-37559 
and  
QUPRODIS, Contract No. IST-2001-38877 and by the Polish Ministry of Scientific
 Research and Information Technology under the (solicited) grant No.
 PBZ-MIN-008/ P03/ 2003.

\bibliography{refbasia}
\end{document}